\documentclass[aps,prd,twocolumn,groupedaddress,nofootinbib,amssymb]{revtex4}

\def\theequation{\arabic{section}.\arabic{equation}}
\usepackage{graphicx,bm,color}
\usepackage{amsmath} 
\usepackage{amssymb} 
\usepackage{amsfonts} 
\usepackage{wrapfig} 
\usepackage{cancel} 
\usepackage{url} 
\usepackage{amsmath} 
\usepackage{amssymb} 
\usepackage{amsfonts} 
\usepackage{graphicx,bm} 
\usepackage{dcolumn} 
\usepackage{color}
\usepackage{color,amsxtra} 
\usepackage{epsf}
\usepackage{amssymb} 
\usepackage{amsmath}
\usepackage{enumerate} 
\usepackage{hhline} 
\usepackage{array} 
\usepackage{tabularx}

\newcommand{\be}{\begin{equation}} 
\newcommand{\ee}{\end{equation}} 
\newcommand{\bea}{\begin{eqnarray}} 
\newcommand{\eea}{\end{eqnarray}} 
\newcommand{\beaa}{\begin{eqnarray*}} 
\newcommand{\eeaa}{\end{eqnarray*}}

\newcommand{\nn}{\nonumber \\} 
\newcommand{\e}{\mathrm{e}}

\allowdisplaybreaks[4]

\begin{document} \tolerance=5000 
\def\theequation{\arabic{section}.\arabic{equation}}

\title{Area-law versus R\'enyi and Tsallis black hole entropies}

\author{Shin'ichi~Nojiri$^{1,2}$, Sergei~D.~Odintsov$^{3,4,5,6}$, and 
Valerio~Faraoni$^{7}$ } 
\affiliation{ $^1$Department of Physics, Nagoya University, 
Nagoya 464-8602, Japan\\ 
$^2$Kobayashi-Maskawa Institute for 
the Origin of Particles and the Universe, Nagoya University, Nagoya 
464-8602, Japan\\ 
$^3$Instituci\'{o} Catalana de Recerca i Estudis 
Avan\c{c}ats (ICREA), Passeig Llu\'{i}s Companys, 23, 08010 Barcelona, 
Spain\\ 
$^4$ Institute of Space Sciences (IEEC-CSIC), C.Can Magrans s/n , 08093 Barcelona, Spain \\
$^5$ Dept. of Physics, Kazan Federal University, Kazan 420008, Russia \\ 
$^6$ Int.~Lab.~Theor.~Cosmology, Tomsk State University of Control Systems and Radioelectronics(TUSUR), 634050 
Tomsk, Russia \\
$^7$ Department of Physics \& Astronomy, 
Bishop's University, 2600 College Street, Sherbrooke, Qu\'ebec, Canada J1M 
1Z7}

\begin{abstract}

The R\'enyi and Tsallis entropies are discussed as possible alternatives 
to the Bekenstein-Hawking area-law entropy. It is pointed out how 
replacing the entropy notion, but not the Hawking temperature and the 
thermodynamical energy may render the whole black hole thermodynamics 
inconsistent. The possibility to relate the R\'enyi and Tsallis entropies 
with the quantum gravity corrected Bekenstein-Hawking entropy is 
discussed. 

\end{abstract} 
\maketitle

\section{Introduction}
\label{sec:1}

Black hole thermodynamics is one of the most interesting recent 
discoveries of theoretical physics. Bekenstein \cite{Bekenstein:1973ur} 
argued that the area $A$ of a black hole horizon has the properties of the 
entropy $\mathcal{S}$ in ordinary thermodynamics and must be proportional 
to it (the proportionality factor was determined later). However, the 
similarity between black hole area and thermodynamical entropy did not 
make sense initially because it was believed that black holes are cold 
objects, until Hawking discovered that the Schwarzschild black hole emits 
quantum radiation with a blackbody spectrum at temperature $T_\mathrm{H}= 
\frac{1}{8\pi G M}$ (now called the Hawking temperature), where $M$ is the 
Schwarzschild mass of the black hole \cite{Hawking:1975vcx}. Here we use 
geometrized units in which the speed of light $c$, the Boltzmann constant 
$K_B$, and the reduced Planck constant $\hbar$ are unity.

Hawking's discovery of black hole radiation originates from the 
application of quantum field theory to curved spacetime and, by extension, 
it implies that all black holes must radiate, making them thermal objects 
and completing Bekenstein's suggestion. The Bekenstein-Hawking entropy 
$\mathcal{S}= A/4$ and the Hawking temperature $T_\mathrm{H}$ allow for 
the construction of a self-consistent black hole thermodynamics 
(\cite{Bardeen:1973gs}, see \cite{Wald:1999vt,Wald,Carlip:2014pma} for 
reviews), which is now an important part of modern theoretical physics. 
Adding a cosmological constant, as in the Schwarzschild-de Sitter/Kottler 
and Schwarzschild-Anti-de Sitter black holes, adds richness to the 
thermodynamical behaviour of black holes. In addition to creating multiple 
horizons which could be viewed as thermodynamical sub-systems, it leads to 
the possibility of the Hawking-Page phase transition 
\cite{Hawking:1982dh}, which was later interpreted in the context of the 
AdS/CFT correspondence as the counterpart of the deconfinement transitions 
for the conformal field theory living on the Anti-de Sitter boundary 
\cite{Maldacena:1997re, Witten:1998zw}. The negative cosmological constant 
turns out to play the role of the pressure, to be added to the 
thermodynamical picture \cite{Kastor:2009wy}, thus extending the structure 
of the phase space and making phase transitions possible. In this context, 
a rich literature on ``black hole chemistry'' has emerged and black hole 
thermodynamics has taken a new lease on life ({\em e.g.}, 
\cite{Kubiznak:2014zwa, Karch:2015rpa, Kubiznak:2015bya, Kubiznak:2016qmn, 
Sinamuli:2017rhp, Tzikas:2018cvs, Astefanesei:2019ehu, Mir:2019ecg, 
Kumar:2020cve} and references therein).

Originally, it came as a surprise that the Bekenstein-Hawking entropy is 
proportional to the black hole area and not to its volume, as in ordinary 
thermodynamics where entropy is an extensive quantity proportional to the 
mass, and then to the volume, of a system. This feature largely remains a 
mystery \cite{Tsallis:2012js}. Recent literature has discussed the 
possibility of replacing the Bekenstein-Hawking entropy with other entropy 
notions based on non-extensive statistics \cite{Bialas:2008fa, 
Tsallis:2012js, Huang:2014pda, Brustein:2014iha, Nishioka:2014mwa, 
Czinner:2015eyk, Dong:2016fnf, Wen:2016itq, Czinner:2017tjq, 
Qolibikloo:2018wqw, Johnson:2018bma, Tannukij:2020njz, Promsiri:2020jga, 
Samart:2020klx, Ren:2020djc, Mejrhit:2020dpo, Nakarachinda:2021jxd, 
Abreu:2021avp}, such as the R\'enyi \cite{Renyi60} and Tsallis 
\cite{Tsallis:1987eu} entropies. However, changing the entropy notion is 
risky because entropy enters many thermodynamical equations and other 
quantities need to be modified in order to keep the whole construction of 
thermodynamics self-consistent. While modifying entropy is challenging and 
many authors have focused on this task, here we point out the risks 
inherent in these modifications for related thermodynamics.  In 
particular, it is problematic to modify the Hawking temperature of 
blackbody radiation based only on non-extensive statistics, while there 
are independent arguments pointing toward the ``correct'' choice of 
thermodynamical energy and black hole mass. When all these aspects are 
considered together, it turns out to be quite difficult to replace the 
Bekenstein-Hawking entropy based solely on the idea of non-extensive 
statistics while keeping black hole thermodynamics self-consistent.

In this work we follow the notation of Ref.~\cite{Wald}: the metric 
signature is ${-}{+}{+}{+}$ and the units are such that the speed of light 
$c$, the Boltzmann constant $K_B$, and the reduced Planck constant $\hbar$ 
are unity.

\section{Standard black hole thermodynamics}
\label{sec:2}

The geometry of the Schwarzschild black hole is described by the line element 
\cite{Wald}
\begin{equation}
\label{dS3BB}
ds^2= - f(r) dt^2 + \frac{dr^2}{f(r)} + r^2 d\Omega^2_{(2)}\, , \quad\quad 
f(r) \equiv 1 - \frac{2GM}{r} \,,
\end{equation}
where $G$ is Newton's constant, $M$ is the black hole mass, 
and $d\Omega^2_{(2)}=d\vartheta^2 +\sin^2 \vartheta \, d\varphi^2$   
is the line element on  the unit two-sphere.  The black hole event horizon is 
located at 
the Schwarzschild radius 
\begin{equation}
\label{horizonradius}
r_\mathrm{H}=2GM\, .
\end{equation}
By considering quantum field theory on the spacetime with this horizon, 
Hawking discovered that the Schwarzschild black hole radiates with a 
blackbody spectrum at the temperature \cite{Hawking:1975vcx}  
\begin{equation}
\label{dS6BB}
T_\mathrm{H} = \frac{1}{8\pi GM}\,.
\end{equation}
The Hawking temperature can be also understood geometrically. 
When $r\sim r_\mathrm{H}$, we define $\delta r$ by $r \equiv r_\mathrm{H} 
+ \delta r$.  Then by Wick-rotating the time coordinate  $t\to i\tau$, the 
line element~(\ref{dS3BB}) is recast as 
\begin{equation}
\label{TH1}
ds^2 \simeq \frac{\delta r}{r_\mathrm{H}} \, d\tau^2
+ \frac{r_\mathrm{H}}{\delta r} \, d(\delta r)^2 + r_\mathrm{H}^2 
\, d\Omega^2_{(2)}\, .
\end{equation}
We further define a new radial coordinate $\rho$ by 
$ d\rho = d \left( \delta r\right)  \sqrt{ r_\mathrm{H}/ \delta r} $, 
that is 
\begin{equation}
\label{TH2}
\rho = 2 \sqrt{ r_\mathrm{H} \delta r} \quad\quad
\mbox{or} \quad\quad \delta r= \frac{\rho^2}{4 r_\mathrm{H}} \,, 
\end{equation}
in terms of which the line element~(\ref{TH1}) becomes  
\begin{equation}
\label{TH3}
ds^2 \simeq \frac{\rho^2}{ 4 r_\mathrm{H}^2} \, \rho^2d\tau^2
+ d\rho^2 + r_\mathrm{H}^2 \, d\Omega_{(2)}^2 \, .
\end{equation}
In order to avoid the conical singularity in the Wick-rotated Euclidean 
space around $\rho\sim 0$, 
we require the periodicity of the Euclidean time coordinate $\tau$ 
\begin{equation}
\label{TH4}
\frac{\tau}{ 2 r_\mathrm{H}} 
\sim \frac{\tau}{ 2 r_\mathrm{H}} + 2\pi \,.
\end{equation}
Because the inverse of the period $t_0$ of the Euclidean time coordinate  
corresponds to the temperature, as in the Euclidean path integral 
formulation of the finite temperature field theory for any field $\phi$, 
\begin{equation}
\label{TH5}
\int \left[ D\phi \right] \e^{ \, \int_0^{t_0} dt \, L(\phi)}
= \mathrm{Tr}\left( \, \e^{-t_0 H} \right) = \mathrm{Tr}\left( \, \e^{ - 
\frac{H}{T} } \right) \, ,
\end{equation}
one finds that the Schwarzschild black hole has temperature $T$, 
which is nothing but the Hawking temperature~(\ref{dS6BB}), 
\begin{equation}
\label{TH6}
T = \frac{1}{4\pi r_\mathrm{H}}= \frac{1}{8\pi GM} \equiv T_\mathrm{H} \,.
\end{equation}
Thus, the Hawking temperature can be obtained solely from the geometry of 
the spacetime endowed with the event horizon. 

We can derive also the entropy $\mathcal{S}$ from the geometrical point 
of view. Eq.~(\ref{TH5}) tells us that the partition function 
$Z(T)$ and the 
free energy $F(T)$ are given by
\begin{equation}
\label{FZ}
\e^{- \frac{F(T)}{T}} = Z(T) =  \mathrm{Tr}\left(\, \e^{ - 
\frac{H}{T} }\right)  
= \int \left[ D\phi \right] \e^{S(\phi)} \, ,
\end{equation}
with the periodic boundary condition that the Euclidean time has  
period $1/T $.  In Eq.~(\ref{FZ}), $S(\phi)=\int_0^{t_0} dt \, L(\phi)$ is 
the Euclidean action. In the low-temperature regime when $T$ is 
sufficiently small, the path integral~(\ref{FZ}) can be estimated as 
\begin{equation}
\label{FZ2}
\int \left[ D\phi \right] \e^{S(\phi)} \sim \e^{S \left( \phi_\mathrm{cl} 
\right)} 
\end{equation}
in the WKB approximation, where $\phi_\mathrm{cl}$ is a classical solution of the 
field equations  
given by the Euclidean action $S(\phi)$. Then Eq.~(\ref{FZ}) allows us to 
estimate the free energy $F$ which, in turn, gives the entropy 
$\mathcal{S}$ by using the thermodynamical relations. 

In order to estimate the free energy $F$, we consider  the 
Schwarzschild-Anti-de Sitter geometry 
\begin{eqnarray}
\label{SAdS1}
&& ds^2 = - f(r) dt^2 + \frac{dr^2}{f(r)} + r^2 d\Omega_{(2)}^2\,,\\ 
&&\nonumber\\
&& f(r) \equiv 1 - \frac{2GM}{r} + \frac{r^2}{l^2}\,,
\end{eqnarray}
where $\Lambda=-3/l^2$ is the cosmological constant (the reason why we 
consider the Schwarzschild-Anti-de Sitter spacetime 
instead of the Schwarzschild one with the flat Minkowski background 
will be explained later). 

We now rewrite the function $f(r)$ in the form 
\begin{eqnarray}
&& f(r) = \frac{\left(r-r_\mathrm{H} \right) \left( r + 
\frac{r_\mathrm{H}}{2} + i a \right) \left( r + \frac{r_\mathrm{H}}{2} - 
i a \right)}{l^2 r} \,,\nonumber\\
&&\\
&&  r_\mathrm{H}\left( \frac{r_\mathrm{H}^2}{4} + a^2 \right) = 2GM l^2 \, , 
\quad 
 - \frac{3r_\mathrm{H}^2}{4} + a^2 = l^2 \,.\nonumber\\
&& \label{SAdS2}
\end{eqnarray} 
Eliminating $a$ in the last two equations
yields
\begin{equation}
\label{SAdSC1}
r_\mathrm{H} \left( r_\mathrm{H}^2 + l^2 \right) = 2GM l^2 \, .
\end{equation}
When $r\sim r_\mathrm{H}$, $f(r)$ behaves as 
\begin{equation}
\label{SAdS4}
f(r) \sim \frac{\frac{ 9r_\mathrm{H}^2}{4}+a^2}{l^2 r_\mathrm{H}} 
\left(r-r_\mathrm{H} \right) 
\end{equation}
and the Hawking temperature is 
\begin{equation}
\label{SAdS5}
T_\mathrm{H} \simeq \frac{\frac{ 9r_\mathrm{H}^2}{4}+a^2}{4 \pi l^2 
r_\mathrm{H}} 
= \frac{3 r_\mathrm{H}^2 + l^2}{4 \pi l^2 r_\mathrm{H}} 
= \frac{1}{4\pi l^2}\left( 3 r_\mathrm{H} + \frac{l^2}{r_\mathrm{H}} 
\right)  \, ,
\end{equation}
where we have used the last equation~(\ref{SAdS2}) to substitute for 
$a$. 

After the Wick rotation, the action becomes  
\begin{eqnarray}
S &=& \frac{1}{16\pi G} \int d^4 x \sqrt{-g} \left( R + \frac{6}{l^2}  
\right) \nonumber\\
&=& - \frac{3}{2Gl^2} \int_0^{ 1/ T_\mathrm{H} } dt  
\int_{r_\mathrm{H}}^L dr \, r^2 \nonumber\\ 
&=& \frac{r_\mathrm{H}^3 - L^3}{2G\, l^2 T_\mathrm{H}} \,,\label{SAdS3}
\end{eqnarray} 
where we have introduced a cutoff $L$ to regulate the divergence of the 
action~(\ref{SAdS3}).  
It is instructive to consider the difference between the 
action  of the Schwarzschild-Anti-de Sitter spacetime and that of the pure 
Anti-de Sitter spacetime. 
We determine the period of the Euclidean time $1/\tilde 
T_\mathrm{H}$ in Anti-de Sitter space 
so that the physical length (between $t=0$ and $t=1/{\tilde 
T}_\mathrm{H}$)  equals the physical length (between 
$t=0$ and $t= 1/ T_\mathrm{H}$) 
in the Schwarzschild-Anti-de Sitter spacetime: 
\begin{equation}
\label{SAdSB1}
\left( 1 - \frac{2GM}{L} + \frac{L^2}{l^2} \right)^{1/2} 
\frac{1}{T_\mathrm{H}} 
= \left( 1 + \frac{L^2}{l^2} \right)^{1/2} \frac{1}{{\tilde 
T}_\mathrm{H}} \, ,
\end{equation}
or
\begin{equation}
\label{SAdSB2}
\frac{1}{{\tilde T}_\mathrm{H}} = \frac{1}{T_\mathrm{H}} \sqrt{ 1 - 
\frac{2GM}{L\left( 1 + \frac{L^2}{l^2} \right)}} 
\sim \frac{1}{T_\mathrm{H}} \left( 1 - \frac{GMl^2}{L^3} \right) \, .
\end{equation}
Then, the action $S_\mathrm{AdS}$ for Anti-de Sitter spacetime  is 
\begin{eqnarray}
S_\mathrm{AdS} &=& - \frac{3}{2Gl^2} \int_0^{1/{\tilde T}_\mathrm{H} } 
dt \int_0^L dr \, r^2 = - \frac{L^3}{2G l^2\, {\tilde T}_\mathrm{H}} 
\nonumber\\
& \sim & - \frac{L^3}{2G  l^2 T_\mathrm{H}} + \frac{GMl^2}{2G l^2 
T_\mathrm{H}}  \nonumber\\
&=&  - \frac{L^3}{2G l^2 T_\mathrm{H}} + \frac{r_\mathrm{h} \left( 
r_\mathrm{H}^2 + l^2 \right) }{4G l^2 T_\mathrm{H}} \,,\label{SAdS3B}
\end{eqnarray}
where we have used Eq.~(\ref{SAdSC1}). In the limit $L\to \infty$, the 
action $S_\mathrm{BH}$ for the black hole in Anti-de Sitter space reduces to 
\begin{equation}
\label{BH1}
S_\mathrm{BH} = S - S_\mathrm{AdS} = \frac{r_\mathrm{H} \left( 
r_\mathrm{H}^2 - l^2 \right) }{4G \, l^2 T_\mathrm{H}} 
\end{equation}
and the free energy $F$ is 
\begin{equation}
\label{SAdS8}
F=  - T_\mathrm{H} S_\mathrm{BH} = - \frac{r_\mathrm{H} \left( 
r_\mathrm{H}^2 - l^2 \right) }{4G l^2} \, .
\end{equation}
Upon use of the thermodynamical relations 
\begin{equation}
\label{SAdS9}
E = F - T_\mathrm{H}\, \frac{d F}{d T_\mathrm{H}}\, , \quad \quad 
\mathcal{S} = \frac{E - F}{T_\mathrm{H}} = -  \frac{d F}{d T_\mathrm{H}}  
\end{equation}
for the thermodynamical energy $E$ and the entropy $\mathcal{S}$, we find 
the entropy
\begin{align}
\label{SAdS10}
\mathcal{S} =& - \frac{ dF/d r_\mathrm{H} }{ d T_\mathrm{H}/ d 
r_\mathrm{H}}
= \frac{ \frac{3 r_\mathrm{H}^2 - l^2 }{4G l^2}}{ \frac{1}{4\pi l^2}\left( 
3 - \frac{l^2}{r_\mathrm{H}^2} \right) }
= \frac{\pi r_\mathrm{H}^2}{G}=\frac{A}{4G}\, ,
\end{align}
where $A = 4\pi {r_\mathrm{H}}^2$ is the horizon area.  The 
expression~(\ref{SAdS10}) is valid in the limit of a flat Minkowski 
background $l^2\to \infty$, thus we obtain the Bekenstein-Hawking 
entropy from a geometrical viewpoint. We should note, however, that the 
action~(\ref{SAdS3}) vanishes in the Minkowski background $l^2\to \infty$, hence, 
the ``Schwarzschild'' black hole in Anti-de Sitter 
space with finite $l$ somehow plays the role of the regularization of the 
Schwarzschild black hole in flat Minkowski background.

Another consideration is in order: restoring the constants, the 
Hawking temperature and the Bekenstein-Hawking entropy read
\be
T_\mathrm{H}= \frac{\hbar c^3}{8\pi G K_B M} \,,  \quad 
\mathcal{S}= \frac{c^2 A}{4G \hbar} \,,
\ee
then the free energy $F=E-TS $ does not contain the reduced Planck 
constant $\hbar$ and, in this sense, it is a classical quantity like $E$. 
Non-extensive black hole entropies that are inspired by modified 
uncertainly principles 
or quantum gravity corrections should not affect $F$ unless they correct 
$T_\mathrm{H}$. Then, according to 
Eq.~(\ref{SAdS10}), the entropy $\mathcal{S}$ is only affected through 
corrections to the black hole temperature. 
However, if this temperature remains the Hawking 
temperature $T_\mathrm{H}$, $\mathcal{S}$ cannot receive corrections, or else 
also Eq.~(\ref{SAdS10}) must be modified accordingly. 
Although a bit hand-waving,\footnote{In the sense that quantum gravity is 
expected to correct the Hawking temperature at some level.} 
this argument makes 
the point that 
thermodynamics (including black hole thermodynamics) must be a consistent 
theory and modifying this or that thermodynamic quantity {\em ad hoc} 
usually has unwanted consequences for the rest of the theory.

\section{Impossibility of non-area law entropy}
\label{sec:3}

In the previous section, we have summarized the derivation of both  the Hawking 
temperature and the  Bekenstein-Hawking entropy from the viewpoint of the 
geometry. As is  
well known,  the area law for the Bekenstein-Hawking 
entropy \cite{Bekenstein:1973ur} can always be obtained if we identify the 
thermodynamical energy $E$ with the black hole mass $M$, $E=M$, and the 
temperature 
of the system with the Hawking temperature (\ref{dS6BB}) 
\cite{Hawking:1975vcx}, 
$T=T_\mathrm{H}$.  In fact, the thermodynamical relation $ 
dE= T d\mathcal{S} $ yields
\begin{equation}
\label{S1}
d\mathcal{S}=\frac{dE}{T}=8\pi GM dM =d\left( 4 \pi G M^2 \right) \, ,
\end{equation}
that is, 
\begin{equation}
\label{S2}
\mathcal{S}=4 \pi G M^2 + \mathcal{S}_0\,,
\end{equation}
where $S_0$ is an integration  constant. If we assume $\mathcal{S}=0$ 
when $M=0$, that is, in the absence of the black hole, then 
$\mathcal{S}_0=0$ and  
\begin{equation}
\label{S3}
\mathcal{S}= \frac{\pi {r_\mathrm{H}}^2}{G} = \frac{A}{4G}
\end{equation}
where $A\equiv 4\pi {r_\mathrm{H}}^2$ is the horizon area. Thus, we have 
shown that the Bekenstein-Hawking entropy ({\em i.e.}, the area law for 
the black hole entropy) can be obtained by assuming $E=M$ and 
$T=T_\mathrm{H}$ by using the thermodynamical relation 
$d\mathcal{S}=dE/T$.

Two questions arise naturally:

\begin{enumerate}

\item Can we identify the thermodynamical energy $E$ with the black hole 
mass $M$ ({\em i.e.}, $E=M$)?  Furthermore, if the black hole is not 
Schwarzschild nor isolated, there is no Arnowitt-Deser-Misner mass: should 
then $M$ be the quasilocal mass contained in the horizon sphere, or the 
``black hole part'' of it? If so, which quasilocal mass? Several 
quasilocal mass prescriptions exist in the literature (see 
Ref.~\cite{Szabados:2009eka} for a review).

\item Is the temperature of the black hole given by the Hawking 
temperature, $T=T_\mathrm{H}$?

\end{enumerate}

To answer the first question, consider the following {\em gedankenexperiment}: 
assume that there is an infalling spherically symmetric shell of dust with 
mass $M$ and initial radius sufficiently large. By virtue of the Birkhoff theorem 
\cite{Wald}, 
the spacetime outside the shell is the Schwarzschild one~(\ref{dS3BB}), 
where $M$ in (\ref{dS3BB}) is the mass of the shell. Inside the shell, 
spacetime is empty and flat. The shell collapses, its radius becoming 
smaller and smaller. When the shell crosses its Schwarzschild 
radius~(\ref{horizonradius}), a black hole is formed. The geometry is  
always asymptotically flat and the shell mass $M$ appearing in the 
Schwarzschild line element is surely the energy $E$ of the system, $E=M$.  
This energy is conserved during the collapse of the shell because, due to 
the Birkhoff theorem, the geometry outside of it does not change during 
the collapse: the energy of the final black hole must be the mass of the 
shell.\footnote{Due to spherical symmetry, gravitational waves (which are 
quadrupole to lowest order) are not emitted during the collapse and cannot 
carry away energy.} Therefore, the 
thermodynamical energy $E$ is the black hole mass $E=M$.

Regarding the mass concept used, the answer is easy for the 
Schwarzschild black hole: the Schwarzschild mass appearing in the line 
element is the obvious choice. It coincides with the 
Misner-Sharp-Hernandez quasilocal mass $M_\mathrm{MSH}$ defined in any 
spherically 
symmetric spacetime by \cite{Misner:1964je,Hernandez:1966zia}
\be
1-\frac{2GM_\mathrm{MSH} }{R} = \nabla^c R \nabla_c R \,,
\ee
where $R$ is the areal radius (which coincides with $r$ in the 
Schwarzschild case~(\ref{dS3BB})). Other quasilocal masses do not reproduce the 
standard Hawking temperature and the Bekenstein-Hawking entropy. For 
example, consider the Brown-York quasilocal energy \cite{Brown:1992br} as  
a possible candidate to the role of thermodynamical energy. For a spherically 
symmetric metric of the form
\be
ds^2 =-N^2(t,r) dt^2 +\frac{dr^2}{f(t,r)} +r^2 d\Omega_{(2)}^2 \,,
\ee
the Brown-York energy is \cite{Brown:1992br,Szabados:2009eka}
\be
E_\mathrm{BY}= \frac{r}{G} \Big[ 1-f(t,r) \Big] \,.
\ee
For the Schwarzschild metric, the Brown-York energy is 
radius-dependent:
\be
E_\mathrm{BY}= \frac{r}{G} \left(1-\sqrt{1-\frac{2M}{r}} \, \right) \,;
\ee
at the Schwarzschild horizon $r=r_\mathrm{H}$, we have 
$E_\mathrm{BY}=2M$. If we 
identify the thermodynamical energy with the Brown-York energy and the 
black hole temperature with the Hawking temperature, the   
relation $Td\mathcal{S}=dE$ yields 
\be
d\mathcal{S}_\mathrm{BY}= \frac{ dE_\mathrm{BY}}{T_\mathrm{H}} = 16\pi G M dM
\ee
and, integrating,
\be
\mathcal{S}_\mathrm{BY}= 8\pi M^2 =\frac{A}{2G} \,,
\ee
which is unphysical. The only way to reconcile the Brown-York energy 
prescription \cite{Brown:1992br} with the relation $Td\mathcal{S}=dE$ is by 
introducing the 
Brown-York temperature $T_\mathrm{BY}=2T_\mathrm{H}$, which disagrees with 
Hawking's fundamental result \cite{Hawking:1975vcx}. One does not see how (or 
why) Hawking's derivation of the black hole temperature should be modified to make 
it agree with the Brown-York mass prescription in thermodynamics. 

Let us consider now the Schwarzschild-Anti-de Sitter black hole (parallel 
considerations apply to the Schwarzschild-de Sitter case). 
For the Schwarzschild-Anti-de Sitter black hole, the  
Misner-Sharp-Hernandez mass is\footnote{For the Schwarzschild-Anti-de 
Sitter black hole, the Brown-York energy 
gives again an unphysical result, {\em i.e.}, 
$  E_\mathrm{BY}= 2M- \frac{r^3}{G\, l^2} $.} 
\be
M_\mathrm{MSH} = M-\frac{r^2}{2G\, l^2} \,,
\ee
where it is easy to isolate the black hole contribution from the one due 
to the negative cosmological constant, and the Hawking temperature and 
the Bekenstein-Hawking entropy should refer to these contributions (at least 
in the limit of small black holes). This happens also for more general black 
holes  embedded in a cosmological ``background'' described by  
the time-dependent McVittie metric \cite{McVittie} (this geometry contains the  
Schwarschild-de Sitter/Kottler metric as a special case) \cite{Carrera:2009ve, 
Faraoni:2015ula}.  However, this is not the case for other cosmological black 
holes, for example for the Sultana-Dyer black hole \cite{SultanaDyer05} for which 
the Misner-Sharp-Hernandez/Hawking-Hayward mass includes a third term coupling 
the black hole and the cosmological energies \cite{Faraoni:2015ula}. 
However it can  be argued 
that, at least for the Sultana solution, the time-dependence of 
both black hole and cosmological horizons  precludes a discussion with 
equilibrium thermodynamics while, in general,  an adiabatic expansion for 
dynamical 
black holes should recover the Schwarzschild thermodynamics.  

It is interesting 
that the thermodynamics 
of dynamical apparent horizons (as opposed to static null event horizons) 
employs the Hawking-Hayward quasilocal energy as the thermodynamical 
energy of  a black hole, which reduces to the Misner-Sharp-Hernandez mass 
in spherical symmetry \cite{Hayward:1994bu}. Due to the 
time-dependence, the first law of thermodynamics must be generalized to 
include  an energy supply vector 
\cite{Hayward:1998ee, Hayward:1997jp}. In 
spherical symmetry, the Clausius definition of entropy coincides with the 
entropy obtained from Wald's Noether charge method \cite{Wald:1993nt, 
Iyer:1994ys} supplemented by the 
Kodama flow \cite{Hayward:1998ee}, and the Kodama temperature 
(which reduces to $T_\mathrm{H}$ in the static case) 
\cite{Kodama:1979vn}. Now, both 
of these prescriptions give back the Bekenstein-Hawking entropy in the 
static case \cite{Hayward:1998ee,Hayward:1997jp}, which receives support 
from the more involved discussion of dynamical black hole horizons. 
Another indication of the privileged role of the Misner-Sharp-Hernandez 
mass among the spectrum of quasilocal masses available in the literature 
\cite{Szabados:2009eka} comes from 
the fact that it is the Noether charge associated with the conservation of 
the Kodama current \cite{Hayward:1994bu}  (which always holds in spherical 
symmetry  \cite{Kodama:1979vn}).  It is the Misner-Sharp-Hernandez mass 
that is used as thermodynamical energy in the thermodynamics of dynamical 
apparent horizons: this is a self-consistent thermodynamical picture  that 
uses in an essential way the Kodama time, in the absence of a timelike 
Killing 
vector (see Ref.~\cite{Vanzo:2011wq} for a review of the relevant tunneling 
formalism). Replacing the black hole mass used in the thermodynamics of 
static black hole event horizons with another energy (quasilocal or not) 
would imply the same replacement in the thermodynamics of apparent 
horizons and would make it inconsistent.

Regarding the second question about the temperature, note that the Hawking 
radiation is obtained if the geometry with horizon is prescribed and the 
standard Hawking temperature is the parameter appearing in the thermal 
distribution of the emitted Hawking radiation. If we place the black hole 
in a heat bath at temperature $T$, thermal equilibrium between the black 
hole radiation and the heat bath occurs when the radiation temperature 
equals the temperature of the heat bath, $T=T_\mathrm{H}$. Therefore, we 
can use the heat bath as a thermometer. The temperature measured by the 
heat bath must be the standard Hawking temperature of the Hawking 
radiation and, therefore, we identify the latter with the black hole 
temperature.

Entropy constructs other than the Bekenstein-Hawking one  are often 
considered in the literature, for example the R\'enyi entropy 
\cite{Czinner:2015eyk,Tannukij:2020njz,Promsiri:2020jga, Samart:2020klx} 
\begin{equation}
\label{RS1}
\mathcal{S}_\mathrm{R}=\frac{1}{\alpha} \ln \left( 1 + \alpha 
\, \mathcal{S} 
\right) \, ,
\end{equation} 
where $\alpha$ is a parameter  and $\mathcal{S}$ is the 
Bekenstein-Hawking entropy~(\ref{S3}), which is recovered in the limit 
$\alpha\rightarrow 0$. In this case, Eq.~(\ref{S2})  with 
$\mathcal{S}_0=0$ gives
\begin{equation}
\label{RS2}
\mathcal{S}_\mathrm{R}=\frac{1}{\alpha} \ln \left( 1 + 4 \pi \alpha G M^2 
\right) \, .
\end{equation}
If the mass $M$ coincides with the energy $E$ of the system due to  the 
energy conservation  as in
\cite{Czinner:2015eyk,Tannukij:2020njz,Promsiri:2020jga,Samart:2020klx}, 
in order for this system  to be consistent with the thermodynamical 
equation $d\mathcal{S}=dE/T$, one needs to define the ``R\'enyi  
temperature'' $T_\mathrm{R}$ by 
\begin{equation}
\label{TR1}
\frac{1}{T_\mathrm{R}} \equiv \frac{d\mathcal{S}_\mathrm{R}}{dM} = 
\frac{8 \pi G M}{1 + 4 \pi \alpha G M^2} \,,
\end{equation}
that is, 
\begin{equation}
\label{TR2}
T_\mathrm{R} = \frac{1}{8\pi GM} + \frac{\alpha M}{2} = T_\mathrm{H} + 
\frac{\alpha}{16 \pi  G T_\mathrm{H}} 
\end{equation}
which is, of course, quite different from the Hawking temperature 
$T_\mathrm{H}$.  Therefore, this ``R\'enyi temperature'' $T_\mathrm{R}$ is not 
the temperature perceived by any observer detecting Hawking radiation, 
which is the concept that started the research area  of black hole thermodynamics. 
This fact tells us that the ``R\'enyi temperature'' $T_\mathrm{R}$ is likely 
physically irrelevant for black hole thermodynamics. 

In Eq.~(\ref{TR1}), we assumed that the thermodynamical energy $E$ is the 
black hole mass $M$ and we obtained an unphysical result. One wonders what 
the result would be if we assumed that the thermodynamical temperature 
$T$ coincides with the Hawking temperature $T_\mathrm{H}$, instead of 
assuming $E=M$.

Under the assumptions $T=T_\mathrm{H}$ and 
$\mathcal{S}=\mathcal{S}_\mathrm{R}$, using 
$dE=Td\mathcal{S}$ we 
find the corresponding thermodynamical energy $E_\mathrm{R}$: 
\begin{eqnarray}
dE_\mathrm{R} &=& T_\mathrm{H} d\mathcal{S}_\mathrm{R}
= \frac{1}{8\pi GM} \, \frac{8 \pi GM dM}{1 + 4\pi \alpha G M^2}\\
&&\nonumber\\
&=&  \frac{dM}{1 + 4\pi \alpha G M^2} 
\label{RE1}
\end{eqnarray}
and, integrating, 
\begin{equation}
\label{RE2}
E_\mathrm{R} = \frac{\arctan \left(  \sqrt{4\pi \alpha \, G}\, M \right) 
}{ \sqrt{4\pi \alpha \, G}}  
= M - \frac{4\pi \alpha \, GM^2 }{3} + \mathcal{O}\left( \alpha^2 \right) 
\,,
\end{equation}
where we have fixed the integration constant so that $E_\mathrm{R}=0$ when 
$M=0$. Due to the correction $- \frac{4\pi \alpha GM^2 }{3} + 
\mathcal{O}\left( \alpha^2 \right)$, the expression~(\ref{RE2}) of the 
thermodynamical energy $E_\mathrm{R}$ obtained differs from the black hole 
mass $M$, $E_\mathrm{R}\neq E$, in a way that has no obvious physical 
interpretation. More important, if the spherically symmetric 
dust shell collapses to a Schwarzschild black hole, the result obtained 
above seems to conflict with energy conservation.

Another entropy notion encountered frequently in the 
literature and motivated by non-extensive statistics is the Tsallis entropy 
\cite{Tsallis:1987eu}, which suggests as a  potential alternative 
to the Bekenstein-Hawking entropy for black holes the quantity  
\begin{equation}
\label{TS1}
\mathcal{S}_\mathrm{T} = \frac{A_0}{4G} \left( \frac{A}{A_0} 
\right)^\delta 
\end{equation}
instead of the R\'enyi-like entropy~(\ref{RS1})  (\cite{Ren:2020djc},  
see also \cite{Nojiri:2019skr}). Here $A_0$ is a constant with the 
dimensions of 
a length squared and $\delta$ is 
a parameter that quantifies the non-extensivity.  In the case $\delta = 
1$, we obtain the standard Bekenstein-Hawking entropy~(\ref{SAdS10}) 
or~(\ref{S3}). 

In Ref.~\cite{Nojiri:2019skr}, cosmology with the 
Tsallis-type entropy (\ref{TS1}) has been studied. The corresponding 
late-time universe contains an effective dark energy, which could be 
phantom or quintessence, without an effective cosmological constant. One 
obtains an effective cosmological constant from the generalized 
non-extensive Tsallis type entropy also in the inflationary era of the 
early universe. 

As done in Eq.~(\ref{TR1}), assuming that the thermodynamical energy $E$ 
is given by the black hole mass $M$, we find $A=4\pi \left( 2GM \right)^2 
= 
16\pi G^2 E^2$ and the corresponding Tsallis entropy 
\begin{equation}
\label{TS2}
\mathcal{S}_\mathrm{T} = \frac{A_0}{4G} \left( \frac{16\pi G^2 E^2}{A_0} 
\right)^\delta \, .
\end{equation}
We may define also the ``Tsallis temperature'' 
\begin{eqnarray}
T_\mathrm{T} & \equiv & \frac{dE}{d\mathcal{S}_\mathrm{T}}
= \frac{2G}{\delta A_0 E^{2\delta - 1}} \left( \frac{A_0}{16\pi G^2} 
\right)^\delta  \nonumber\\
&=& \frac{2G}{\delta A_0 M^{2\delta - 1}} \left( \frac{A_0}{16\pi G^2} 
\right)^\delta  \label{TS3}
\end{eqnarray}
which is, of course, different from the Hawking temperature~(\ref{TH6})
unless $\delta =1$. Then, instead of identifying the black 
hole mass $M$ with the thermodynamical energy $E$, we could assume that the 
temperature is the Hawking temperature~(\ref{TH6}). Then, since   $ A= 
4\pi \left( 4\pi T_\mathrm{H} \right)^{-2} = \frac{1}{4\pi 
{T_\mathrm{H}}^2}$, we would find

\begin{equation}
\label{TS4}
\mathcal{S}_\mathrm{T} = \frac{{A_0}^{1-\delta}}{4 G \left(4\pi 
{T_\mathrm{H}}^2 \right)^\delta} 
\end{equation}
and then we may define the ``Tsallis energy'' $E_\mathrm{T}$ by 
\begin{equation}
\label{TS5}
d E_\mathrm{T} = T_\mathrm{H} d\mathcal{S}_\mathrm{T} 
= - \frac{\delta {A_0}^{1-\delta} d T_\mathrm{H}}{2 G 
\left(4\pi\right)^\delta  {T_\mathrm{H}}^{2\delta}} 
\end{equation}
or, integrating, 
\begin{equation}
\label{TS6}
E_\mathrm{T} = \frac{\delta {A_0}^{1-\delta} }{2 \left( 2\delta -1 
\right) G \left(4\pi\right)^\delta  {T_\mathrm{H}}^{2\delta-1}} 
= \frac{\delta {A_0}^{1-\delta} \left( 8\pi GM \right)^{2\delta -1} }{2 
\left( 2\delta -1 \right) G \left(4\pi\right)^\delta } \,,
\end{equation}
where we have again fixed the integration constant by imposing 
$E_\mathrm{T}=0 $ at $M=0$. 
The standard relation $E_\mathrm{T} = M$ is reproduced for $\delta=1$, 
but $E_\mathrm{T} \neq M$ otherwise.

In order to build a thermodynamic theory one needs to identify, in 
addition to the entropy,  the thermodynamical energy $E$ and  
the temperature $T$. If 
we choose as entropy any quantity different from the Bekenstein-Hawking 
entropy, then we lack physical grounds and we can choose the 
thermodynamical energy $E$ and/or the temperature $T$ as we wish. Not 
surprisingly, this procedure leads to an unphysical outcome, {\em i.e.},  
it conflicts with energy conservation and/or with the established physics 
of the Hawking radiation process. To conclude, 
the assumed R\'enyi  entropy~(\ref{RS1}) cannot be the black hole entropy, 
although it might 
be the entropy of a system different from the black hole 
(see, for example, Ref.~\cite{Dong:2016fnf}), as in the case of the 
Tsallis entropy. The Tsallis entropy  is obtained 
as a result of the Fermi, Bose, or Boltzmann statistics for the system 
with long-range forces by the standard statistical mechanics procedure of  
using the Hamiltonian and counting the number of states.  In this sense, there is 
no physical counterpart that could be the R\'enyi statistics. 
The R\'enyi entropy could be an index specifying the information, with no 
relation with the statistics of any physical system.

\section{Discussion and conclusions}
\label{sec:4}

The idea of replacing the standard Boltzmann-Gibbs statistics with 
non-extensive statistics has led to the R\'enyi \cite{Renyi60} and Tsallis 
\cite{Tsallis:1987eu} entropies. Since the Bekenstein-Hawking black hole 
entropy is non-extensive, several authors have considered the possibility 
of replacing the Bekenstein-Hawking entropy with the R\'enyi or the 
Tsallis one. This proposal has somehow mixed with the parallel idea of 
correcting the Bekenstein-Hawking entropy with modifications due to the 
Generalized Uncertainty Principle, or with other ideas. We have pointed 
out the fact that changing the entropy goes hand in hand with changing 
other thermodynamical quantities, or else the entire thermodynamical 
theory may become inconsistent, but this creates more problems of 
principle.

For spherical black holes, one could think of reconciling the Hawking 
temperature $T_\mathrm{H}$ with some entropy notion different from the 
Bekenstein-Hawking entropy by adopting as thermodynamical energy some 
quasilocal energy $E_x$, for which various prescriptions can be found in 
the literature \cite{Szabados:2009eka}. We have discussed the Brown-York 
energy as an example. However, already at first sight, this task appears 
very difficult to say the least (and, even if it was logically possible, 
very contrived from the physical point of view). The explicit forms of the 
R\'enyi and Tsallis entropies make it practically impossible to match 
$T_\mathrm{H}$, $E_x$, and $ \mathcal{S}_\mathrm{R}$ (or 
$\mathcal{S}_\mathrm{T}$) while satisfying the relation 
$Td\mathcal{S}_\mathrm{R,T}= dE_x$ even for the simple Schwarzschild black 
hole, which is static and isolated. Indeed, the case for the 
Bekenstein-Hawking entropy seems rather compelling at this point.

Let us examine now possible loopholes to the area law for entropy. In the 
previous sections, we have shown that the area law of the 
Bekenstein-Hawking entropy can always be obtained if we assume that the 
thermodynamical energy $E$ is identified with the mass $M$, $E=M$, and the 
temperature of the system is the Hawking temperature. Moreover, other 
forms of the entropy such as the R\'enyi and the Tsallis entropies lead to 
unphysical results.

In order to show that $E=M$, we have used the argument of the dust shell 
collapse in conjunction with energy conservation. This argument might not 
carry over to theories of modified gravity, especially higher derivative 
theories such as $F(R)$ gravity. In the Einstein gravity, the falling 
shell of the dust constitutes an exact solution of the field equations. In 
higher derivative gravity, the junction conditions between the matter and 
the vacuum at the shell becomes more complex and more restrictive 
\cite{Barrabes:1997kk, Bressange:2000qp, Velay-Vitow:2017odc, 
Berezin:2019maf, Berezin:2020mas, Berezin:2020bvs, Aviles:2019xae, 
Olmo:2020fri, Chu:2021uec, Rosa:2021teg}. Therefore, whether the infalling 
dust shell constitutes an exact solution of the field equations remains to 
be determined and the previous argument may not apply. What, is 
more the gravitational constant $G$ becomes a scalar degree of freedom 
$\phi \sim G^{-1}$, the Brans-Dicke-like scalar field \cite{Brans:1961sx}, 
already in scalar-tensor gravity (which includes $F(R)$ theories as a 
subclass \cite{Sotiriou:2008rp, Nojiri:2010wj, Nojiri:2017ncd}). Then, the 
area law for the entropy becomes $\mathcal{S}= \phi A/4$ and the scalar 
$\phi$ gives a contribution to the differential $d\mathcal{S}$ in 
$Td\mathcal{S}=dE$. While it is true that all vacuum black holes of 
scalar-tensor gravity that are spherical, asymptotically flat, static, and 
sit in a minimum of the potential $V(\phi)$ for $\phi$ reduce to the 
Schwarzschild black hole \cite{Hawking:1972qk, Bekenstein:1996pn, 
Sotiriou:2011dz,Bhattacharya:2015iha}, the discussion becomes more 
complicated already for asymptotically de Sitter black holes.

Even in Einstein gravity, if we include quantum corrections there might be 
modifications to the reasoning above. For example, quantum fluctuations of the 
horizon have been discussed in \cite{Barrow:2020tzx}. The quantum fluctuation may 
effectively increase the area of the horizon, changing the entropy. In the 
arguments following Eq.~(\ref{FZ2}) we have used the WKB approximation, which is 
valid at low temperature. But at high temperature the correction given by the $ 
1/T_\mathrm{H} $ expansion becomes large and there could be a violation of the 
entropy area law. The $ 1/T_\mathrm{H} $ correction may be also regarded as a 
quantum corrections because if we include $\hbar$, the factor $t_0= 1/T_\mathrm{H} 
$ in front of the Hamiltonian $H$ in (\ref{TH5}) becomes $\frac{t_0}{\hbar} 
=\frac{1}{\hbar T_\mathrm{H}}$. Such corrections are considered in 
\cite{Cognola:2005de}. Then, if we assume $E=M$ as in~(\ref{TR1}), the Hawking 
temperature might not be modified and the thermal distribution of the radiation 
might be changed from that of blackbody radiation. Such a correction might appear 
as the shift from $\alpha \rightarrow 0$ in the R\'enyi-type entropy~(\ref{RS1}) or 
the shift from $\delta=1$ in the Tsallis-like entropy~(\ref{TS1}).

In general, entropy corresponds to the number of physical degrees of freedom of a 
thermodynamical system. The renormalization of a quantum theory implies that the 
number of degrees 
of freedom depends on the scale. In standard field theory massive modes 
decouple in the low-energy regime and, therefore, the number of degrees of 
freedom decreases. The situation is more complicated in the case of 
gravity: if gravity is described by string theory, an infinite tower of 
massive modes appears at high temperature, which introduces the upper 
bound to the temperature called Hagedorn temperature. Even from a naive 
point of view, if the spacetime fluctuations become large in the high 
temperature regime, the number of degrees of freedom may increase. If, instead,  
gravity becomes topological, the number of degrees of 
freedom will decrease, which could be consistent with holography. In the 
case of the Bekenstein-Hawking entropy~(\ref{S3}) or~(\ref{SAdS10}), 
because the area is given by $A= \frac{1}{4\pi {T_\mathrm{H}}^2}$, in the 
high temperature regime of large $T_\mathrm{H}$, the entropy decreases. 
This happens, of course, because the smaller black hole has higher Hawking 
temperature. The decrease in the entropy corresponds to the loss of 
physical degrees of freedom due to Hawking radiation.

In the case of the R\'enyi-type entropy~(\ref{RS1}), if the parameter 
$\alpha$ is positive, the deviation from the Bekenstein-Hawking entropy 
$\mathcal{S}$ (\ref{S3}) becomes large in the low-temperature region where 
$\mathcal{S}$ becomes large, which might contradict the above 
speculations in which this deviation becomes large in the high-temperature 
region instead. If $\alpha<0$, we may further modify the 
R\'enyi-like entropy as
\begin{equation}
\label{RS1M1}
\mathcal{S}_\mathrm{R}=\frac{1}{\alpha} \ln \left| 1 + \alpha \mathcal{S} 
\right| \,;
\end{equation}
then there is a singularity when $1 + \alpha \mathcal{S} =0$, or
\begin{equation}
\label{RS1M2}
T_\mathrm{H} = T_\mathrm{c} \equiv \sqrt{ - \frac{\alpha}{16\pi G}} \, .
\end{equation}

If we start at temperature higher than $T_\mathrm{c}$, the critical 
temperature $T_\mathrm{c}$ gives a lower bound on the Hawking 
temperature. If, instead, we start at temperature lower than 
$T_\mathrm{c}$, then $T_\mathrm{c}$ sets an upper bound that might 
correspond to the Hagedorn temperature. Therefore, it might be interesting  
to consider a model in which $\alpha$ depends on the temperature and 
becomes negative at high temperature. In the case of 
the Tsallis-like entropy (\ref{TS1}), if the parameter $\delta$ is larger 
than unity, $\delta > 1$, the Tsallis entropy~(\ref{TS1}) becomes 
smaller than the Bekenstein-Hawking entropy at high temperature. If 
$\delta < 1$, the Tsallis type entropy becomes larger than the 
Bekenstein-Hawking entropy at high temperatures. Then, it might be 
interesting to consider a model in which the parameter $\delta$ depends 
on the temperature. If $\delta>1$ at high temperatures, this model might 
correspond to the model becoming topological at  
high $T$. If, instead, $\delta<1$ in the  high-temperature regime, the 
model might correspond to a violent fluctuation of spacetime or to  
string theory.

If the R\'enyi entropy~(\ref{RS1}) comes from a quantum correction, we 
may express its parameter as $\alpha=\alpha_0 \hbar$ and expand 
Eq.~(\ref{RS1}) with respect to $\hbar$ as
\begin{equation}
\label{RS1hbar}
\mathcal{S}_\mathrm{R}=\frac{1}{\alpha_0 \hbar} \ln \left( 1 + \alpha_0 
\hbar \mathcal{S} \right) 
= \mathcal{S} - \frac{\alpha_0 \hbar}{2} \, \mathcal{S}^2 + 
\frac{\alpha_0^2 \hbar^2}{3} \, \mathcal{S}^3 
+ \cdots 
\end{equation}
If $\alpha$ is positive, the leading correction is negative, which could  hint  
at gravity becoming topological by quantum effects, whereas if $\alpha<0$  the 
leading correction is positive, possibly saying that the number of degrees of 
freedom increases as in string theory.  Similarly, by writing $\delta = 1 + 
\delta_0 \hbar$, 
we may expand the Tsallis entropy~(\ref{TS1}) as  
\begin{eqnarray}
\mathcal{S}_\mathrm{T} &=& \frac{A_0}{4G} \left( \frac{A}{A_0} \right)^{1 
+ \delta_0 \hbar} \nonumber\\
&=& \frac{A}{4G} \left[ 1 + \delta_0 \hbar \ln \frac{A}{A_0} + 
\frac{\delta_0^2 \hbar^2}{2!} \left( \ln \frac{A}{A_0} \right)^2 
+ \cdots \right] \,. \nonumber\\
&& \label{TS1hbar}
\end{eqnarray}
Then, if $\delta>0$ the number of degrees of freedom increases for  
large black 
holes where $A>A_0$ and decreases for small black holes with $A<A_0$. 
If $\delta<0$, the number of degrees of freedom increases for small 
black holes 
and decreases for large ones.  We should also note that the logarithmic 
correction in Eq.~(\ref{TS1hbar}) often appears at the first loop in 
quantum field theory.

Another possible loophole might be the case in which there are two 
horizons, as discussed in Ref.~\cite{Volovik:2021upi}, where the entropies 
of the Reissner-Nordstr\"om and of the Kerr black holes have 
been 
investigated, and it is claimed that there might be contributions to 
the entropy from the correlation of the two horizons. As another example, 
we  may consider the Schwarzschild-de Sitter/Kottler spacetime, where two 
horizons (the black hole and the cosmological horizon) appear. The  line 
element is 
\begin{eqnarray}
&&ds^2= - f(r) dt^2 + \frac{dr^2}{f(r)} + r^2 d\Omega_{(2)}^2\, ,\\
&& f(r) = 1 - \frac{2GM}{r} - \frac{r^2}{l^2}\,; \label{SdS1}
\end{eqnarray}
if we assume $l^2>0$, we can rewrite $f(r)$ as  
\begin{eqnarray}
&& f(r) = - \frac{\left(r-r_-\right) \left( r - r_+ \right) \left( r + r_+ 
+ r_- \right)}{l^2 r} \, ,\label{SdS2a}\\
&&\nonumber\\ 
&& r_+ r_- \left( r_+ + r_- \right) = 2GM l^2 \,,\label{SdS2b}\\ 
&&\nonumber\\
&& r_+^2 + r_-^2 + r_+  r_- =l^2 \,,\label{SdS2c}
\end{eqnarray}
where we assume $r_+>r_-$. Then, $r_+$ and $r_-$  are the radii of the 
cosmological and black hole horizons, respectively.  
When $r\sim r_-$, $f(r)$ behaves as 
\begin{equation}
f(r) \sim \frac{\left( r_+ - r_- \right) \left( 2 r_- + r_+ \right)}{l^2 
r_-} \left(r-r_-\right) \, ,\label{SdS4}
\end{equation}
and the Hawking temperature $T_\mathrm{H}^\mathrm{(bh)}$ of the black hole 
horizon is  
\begin{eqnarray}
T_\mathrm{H}^\mathrm{(bh)} &=& \frac{\left( r_+ - r_- \right) \left( 2 r_- 
+ r_+ \right)}{4\pi l^2 r_-} \nonumber\\
&=& \frac{1}{4\pi l^2}\frac{ - 2 r_-^2 + r_- r_+ + r_+^2}{r_-}
= \frac{1}{4\pi l^2}\frac{ l^2 - 3 r_-^2 }{r_-}  \nonumber\\
&=& \frac{1}{4\pi l^2}\left(\frac{ l^2}{r_-} - 3 r_- \right) \,. 
\label{SdS5}
\end{eqnarray}
Similarly, the Gibbons-Hawking temperature of the 
cosmological horizon is \cite{Gibbons:1977mu}
\begin{equation}
\label{SdS5B}
T_\mathrm{H}^\mathrm{(c)} = \frac{\left( r_+ - r_- \right) \left( r_- + 2 
r_+ \right)}{4\pi l^2 r_+} 
\end{equation}
and we should note that 
\begin{equation}
\label{SdS5C}
T_\mathrm{H}^\mathrm{(bh)} - T_\mathrm{H}^\mathrm{(c)} 
= \frac{ \left( r_+ - r_- \right)^2 \left( r_+ + r_- \right) }{4\pi l^2 
r_+ r_- } \geq 0 \,,
\end{equation}
which implies that there could be heat flow from the black hole 
to the cosmological horizon.
 
In Eq.~(\ref{SdS5C}), the equality ``$=$'' holds if and only if  
$r_+=r_-$, 
that is, in the extremal case corresponding to the Nariai spacetime. 
Because $T_\mathrm{H}^\mathrm{(bh)} \neq T_\mathrm{H}^\mathrm{(c)}$ in 
general, even if we Wick-rotate the spacetime into the Euclidean 
signature, we cannot remove both conical singularities 
corresponding to the black hole and the cosmological horizons and  the calculation 
of the free energy performed for the Schwarzschild-Anti-de Sitter spacetime could 
not be applied. In the 
following, we consider the case in which only the conical singularity of 
the  black hole horizon is removed by assuming the period of the 
Euclidean time  $ 1/T_\mathrm{H}^\mathrm{(bh)} $, 
but the conical singularity of the cosmological horizon remains.

The equations~(\ref{SdS2b}), (\ref{SdS2c}) tell us that
\begin{eqnarray}
\label{SdS6}
&& r_- \left( 2 r_+ + r_- \right) \frac{dr_+}{dM} + r_+ \left( r_+ + 2 r_- 
\right) \frac{dr_-}{dM} = 2G l^2 \, ,\nonumber\\
&&\\
&& \left( 2 r_+ + r_- \right) \frac{dr_+}{dM} + \left( r_+ + 2 r_- \right) 
\frac{dr_-}{dM} = 0 \, ,
\end{eqnarray} 
which give
\begin{eqnarray}
\frac{dr_+}{dM}  &=& - \frac{2Gl^2}{\left( r_- - r_+ \right) \left( 2r_+ + 
r_- \right)} \, ,\\
&&\nonumber\\
\frac{dr_-}{dM} &=& - \frac{2Gl^2}{\left( r_+ - r_- \right) \left( r_+ + 2 
r_- \right)}  = - \frac{2Gl^2}{r_+^2 + r_+ r_-  - 2 r_-^2 }\nonumber\\
&&\nonumber\\ 
&=& - \frac{2Gl^2}{l^2 - 3 r_-^2 } \, .
\label{SdS7}
\end{eqnarray}
Then the action is  
\begin{align}
\label{SdS3}
S=&\frac{1}{16\pi G} \int d^4 x \sqrt{-g} \left( R - \frac{6}{l^2}  
\right) \\
&\nonumber\\
= & \frac{3}{2Gl^2} \int_0^{ 1/ T_\mathrm{H}^\mathrm{(bh)} } dt 
\int_{r_-}^{r^+} dr \, r^2 \nn
&\nonumber\\
= & \frac{r_+^3 - r_-^3}{2G l^2 T_\mathrm{H}} 
= \frac{ \left( r_+ - r_-\right) \left( r_+^2 + r_-^2 + r_+ r_- 
\right)}{2G l^2 T_\mathrm{H}^\mathrm{bh}} 
= \frac{ r_+ - r_-}{2GT_\mathrm{H}^\mathrm{bh}}  
\end{align}
and the free energy $F$ reads
\begin{equation}
\label{SdS8}
F= T_\mathrm{H}^\mathrm{(bh)} S = \frac{r_+ - r_-}{2G} \, .
\end{equation}
By using the thermodynamical relations for the thermodynamical 
energy $E$ and the entropy $\mathcal{S}$ in~(\ref{SAdS8}), we then find 
\begin{align}
\label{SdS10}
\mathcal{S} =& - \frac{1}{2G} \frac{ 
 - \frac{2Gl^2}{\left( r_- - r_+ \right) \left( 2r_+ + r_- \right)} + 
\frac{2Gl^2}{\left( r_+ - r_- \right) \left( r_+ + 2 r_- \right)} }
{- \frac{1}{4\pi l^2}\left(- \frac{ l^2}{r_-^2} - 3 \right) 
\frac{2Gl^2}{l^2 - 3 r_-^2 }} \nn
&\nonumber\\
= & - \frac{2 \pi l^2 r_-^2}{G} \frac{ 2 r_- + r_+ + 2 r_+ + r_- }{\left( 
r_+ - r_- \right) \left( r_+ + 2 r_- \right) \left( 2r_+ + r_- \right)} 
\frac{ l^2 - 3 r_-^2}{l^2 + 3 r_-^2} \nn
&\nonumber\\
= & - \frac{6 \pi l^2 r_-^2}{G} \frac{r_+ + r_- }{\left( 2r_+ + r_- 
\right) \left(l^2 + 3 r_-^2 \right)}\, , 
\end{align}
which is negative and, as  a result, the quantity $\mathcal{S}$ given by  
Eq.~(\ref{SdS10}) cannot be 
identified with the entropy. We cannot obtain the area law 
entropy for the Schwarzschild-de Sitter spacetime, which might be 
due to the correlation between the two horizons. 

A possible way to avoid the above problem of the Schwarzschild-de Sitter 
spacetime could be to analytically continue the results~(\ref{SAdS8}) 
and~(\ref{SAdS10}) to the Schwarzschild-Anti-de Sitter spacetime by 
replacing $l^2\to  - l^2$, obtaining
\begin{equation}
\label{SdS8B}
F=- \frac{r_\mathrm{H} \left( {r_-}^2 + l^2 \right) }{4G l^2} \, , 
\quad \quad 
\mathcal{S} = \frac{ - \frac{3 {r_-}^2 + l^2 }{4G l^2}}{ - \frac{1}{4\pi 
l^2}\left( 3 + \frac{l^2}{{r_-}^2} \right) }
=\frac{A}{4G}\, ,
\end{equation}
which could be the standard result.  The differences between 
Eqs.~(\ref{SdS10}) and~(\ref{SdS8B}) originate from the integration region 
of the action, as we find by comparing Eqs.~(\ref{SAdS3})  
and~(\ref{SdS3}) and also come from the subtraction of the background 
in~(\ref{BH1}). This fact tells us that, if we can use the analytic 
continuation $l^2\to - l^2$, the domain of integration should include the 
region outside the cosmological horizon $L\gg r_+$, which could induce a 
conceptual problem in black hole physics. Anyway, the existence of two 
horizons makes it difficult to judge what could be the correct 
prescription to evaluate the entropy. 

In \cite{Nojiri:2019skr}, the holographic dark energy model based on the 
Tsallis-type entropy or its generalization associated with the cosmological horizon was 
proposed. We may consider the creation of a black hole in the holographic dark energy. If 
the holographic dark energy has a higher temperature than the black hole temperature, there 
could occur an energy flow to the black hole and the black hole may consequently grow, which 
may signal an instability of the holographic dark energy. Black hole seeds may 
be generated by fluctuations of the energy density, which could be the density of 
the holographic dark energy. If we regard the holographic dark energy as a kind of 
effective perfect fluid, however, it 
should have a negative effective pressure. This negative pressure generates a  
repulsion in the fluid, which protects a fluctuation from growing and becoming a black 
hole.  In this sense, the collapse of the holographic dark energy to black holes could be 
avoided and the holographic dark energy could be stable. 

The flow of heat energy onto a black hole, if it occurs due to the temperature difference 
between black hole and surrounding dark energy, is difficult to model. 
Ref.~\cite{Babichev:2004yx} models the spherical accretion of phantom dark energy onto a 
black hole in the test fluid approximation but, realistically, the backreaction is not 
described and the same will be true for radial heat flow. It may be possible to describe the 
net heat transferred in an event occurring between two stationary phases, but this would 
probably require a cosmological evolution that is completely {\em ad hoc.}

To conclude, the entropy area law is physically well motivated, unlike its 
R\'enyi and Tsallis potential competitors, and solid motivations seem to 
be needed before departures from it can be taken too seriously. In 
particular, replacing the Bekenstein-Hawking entropy with the R\'enyi,  
Tsallis, or another entropy is premature and rather arbitrary at the 
current stage of knowledge.

\section*{Acknowledgments} 

This work was also partially supported by the Kazan Federal University 
Strategic Academic Leadership Program 
(S.~D.~O), by JSPS Grant-in-Aid for Scientific 
Research (C) No. 18K03615 (S.~N.), and by the Natural Sciences \& 
Engineering Research Council of Canada, Grant No. 2016-03803 (V.~F.).


\begin{thebibliography}{99}


\bibitem{Bekenstein:1973ur}
J.~D.~Bekenstein,
Phys. Rev. D \textbf{7} (1973), 2333-2346
doi:10.1103/PhysRevD.7.2333

\bibitem{Hawking:1975vcx}
S.~W.~Hawking,
Commun. Math. Phys. \textbf{43} (1975), 199-220
[erratum: Commun. Math. Phys. \textbf{46} (1976), 206]
doi:10.1007/BF02345020

\bibitem{Bardeen:1973gs}
J.~M.~Bardeen, B.~Carter and S.~W.~Hawking,
Commun. Math. Phys. \textbf{31}, 161-170 (1973)
doi:10.1007/BF01645742

\bibitem{Wald:1999vt} R.~M.~Wald,
Living Rev. Rel. \textbf{4} (2001), 6 doi:10.12942/lrr-2001-6 
[arXiv:gr-qc/9912119 [gr-qc]].

\bibitem{Wald} R.~M. Wald, {\em General Relativity} (Chicago University 
Press, Chicago, 1984).

\bibitem{Carlip:2014pma}
S.~Carlip,
Int. J. Mod. Phys. D \textbf{23}, 1430023 (2014)
doi:10.1142/S0218271814300237
[arXiv:1410.1486 [gr-qc]].

\bibitem{Hawking:1982dh}
S.~W.~Hawking and D.~N.~Page,
Commun. Math. Phys. \textbf{87}, 577 (1983)
doi:10.1007/BF01208266

\bibitem{Maldacena:1997re}
J.~M.~Maldacena,
Adv. Theor. Math. Phys. \textbf{2}, 231-252 (1998)
doi:10.1023/A:1026654312961
[arXiv:hep-th/9711200 [hep-th]].

\bibitem{Witten:1998zw}
E.~Witten,
Adv. Theor. Math. Phys. \textbf{2}, 505-532 (1998)
doi:10.4310/ATMP.1998.v2.n3.a3
[arXiv:hep-th/9803131 [hep-th]].

\bibitem{Kastor:2009wy}
D.~Kastor, S.~Ray and J.~Traschen,
Class. Quant. Grav. \textbf{26}, 195011 (2009)
doi:10.1088/0264-9381/26/19/195011
[arXiv:0904.2765 [hep-th]].

\bibitem{Kubiznak:2014zwa}
D.~Kubiznak and R.~B.~Mann,
Can. J. Phys. \textbf{93}, no.9, 999-1002 (2015)
doi:10.1139/cjp-2014-0465
[arXiv:1404.2126 [gr-qc]].

\bibitem{Karch:2015rpa}
A.~Karch and B.~Robinson,
JHEP \textbf{12}, 073 (2015)
doi:10.1007/JHEP12(2015)073
[arXiv:1510.02472 [hep-th]].

\bibitem{Kubiznak:2015bya}
D.~Kubiznak and F.~Simovic,
Class. Quant. Grav. \textbf{33}, no.24, 245001 (2016)
doi:10.1088/0264-9381/33/24/245001
[arXiv:1507.08630 [hep-th]].

\bibitem{Kubiznak:2016qmn}
D.~Kubiznak, R.~B.~Mann and M.~Teo,
Class. Quant. Grav. \textbf{34}, no.6, 063001 (2017)
doi:10.1088/1361-6382/aa5c69
[arXiv:1608.06147 [hep-th]].

\bibitem{Sinamuli:2017rhp}
M.~Sinamuli and R.~B.~Mann,
Phys. Rev. D \textbf{96}, no.8, 086008 (2017)
doi:10.1103/PhysRevD.96.086008
[arXiv:1706.04259 [hep-th]].

\bibitem{Tzikas:2018cvs}
A.~G.~Tzikas,
Phys. Lett. B \textbf{788}, 219-224 (2019)
doi:10.1016/j.physletb.2018.11.036
[arXiv:1811.01104 [gr-qc]].

\bibitem{Astefanesei:2019ehu}
D.~Astefanesei, R.~B.~Mann and R.~Rojas,
JHEP \textbf{11}, 043 (2019)
doi:10.1007/JHEP11(2019)043
[arXiv:1907.08636 [hep-th]].

\bibitem{Mir:2019ecg}
M.~Mir, R.~A.~Hennigar, J.~Ahmed and R.~B.~Mann,
JHEP \textbf{08}, 068 (2019)
doi:10.1007/JHEP08(2019)068
[arXiv:1902.02005 [hep-th]].

\bibitem{Kumar:2020cve}
A.~Kumar, S.~G.~Ghosh and S.~D.~Maharaj,
Phys. Dark Univ. \textbf{30}, 100634 (2020)
doi:10.1016/j.dark.2020.100634
[arXiv:2106.15925 [gr-qc]].

\bibitem{Bialas:2008fa}
A.~Bialas and W.~Czyz,
EPL \textbf{83} (2008) no.6, 60009
doi:10.1209/0295-5075/83/60009
[arXiv:0801.4645 [gr-qc]].

\bibitem{Tsallis:2012js}
C.~Tsallis and L.~J.~L.~Cirto,
Eur. Phys. J. C \textbf{73} (2013), 2487
doi:10.1140/epjc/s10052-013-2487-6
[arXiv:1202.2154 [cond-mat.stat-mech]].

\bibitem{Huang:2014pda}
X.~Huang and Y.~Zhou,
JHEP \textbf{02} (2015), 068
doi:10.1007/JHEP02(2015)068
[arXiv:1408.3393 [hep-th]].

\bibitem{Brustein:2014iha}
R.~Brustein and A.~J.~M.~Medved,
Phys. Rev. D \textbf{91} (2015) no.8, 084062
doi:10.1103/PhysRevD.91.084062
[arXiv:1407.4914 [hep-th]].

\bibitem{Nishioka:2014mwa}
T.~Nishioka,
JHEP \textbf{07} (2014), 061
doi:10.1007/JHEP07(2014)061
[arXiv:1401.6764 [hep-th]].

\bibitem{Czinner:2015eyk}
V.~G.~Czinner and H.~Iguchi,
Phys. Lett. B \textbf{752} (2016), 306-310
doi:10.1016/j.physletb.2015.11.061
[arXiv:1511.06963 [gr-qc]].

\bibitem{Dong:2016fnf}
X.~Dong,
Nature Commun. \textbf{7} (2016), 12472
doi:10.1038/ncomms12472
[arXiv:1601.06788 [hep-th]].

\bibitem{Wen:2016itq}
W.~Y.~Wen,
Int. J. Mod. Phys. D \textbf{26} (2017) no.10, 1750106
doi:10.1142/S0218271817501061
[arXiv:1602.08848 [gr-qc]].

\bibitem{Czinner:2017tjq}
V.~G.~Czinner and H.~Iguchi,
Eur. Phys. J. C \textbf{77}, no.12, 892 (2017)
doi:10.1140/epjc/s10052-017-5453-x
[arXiv:1702.05341 [gr-qc]].

\bibitem{Johnson:2018bma}
C.~V.~Johnson,
Int. J. Mod. Phys. D \textbf{28} (2019) no.07, 1950091
doi:10.1142/S0218271819500913
[arXiv:1807.09215 [hep-th]].

\bibitem{Qolibikloo:2018wqw}
S.~Qolibikloo and A.~Ghodsi,
Eur. Phys. J. C \textbf{79} (2019) no.5, 406
doi:10.1140/epjc/s10052-019-6927-9
[arXiv:1811.04980 [hep-th]].

\bibitem{Tannukij:2020njz}
L.~Tannukij, P.~Wongjun, E.~Hirunsirisawat, T.~Deesuwan and C.~Promsiri,
Eur. Phys. J. Plus \textbf{135} (2020) no.6, 500
doi:10.1140/epjp/s13360-020-00517-2
[arXiv:2002.00377 [gr-qc]].

\bibitem{Promsiri:2020jga}
C.~Promsiri, E.~Hirunsirisawat and W.~Liewrian,
Phys. Rev. D \textbf{102} (2020) no.6, 064014
doi:10.1103/PhysRevD.102.064014
[arXiv:2003.12986 [hep-th]].

\bibitem{Samart:2020klx}
D.~Samart and P.~Channuie,
[arXiv:2012.14828 [hep-th]].

\bibitem{Ren:2020djc}
J.~Ren,
JHEP \textbf{05} (2021), 080
doi:10.1007/JHEP05(2021)080
[arXiv:2012.12892 [hep-th]].

\bibitem{Mejrhit:2020dpo}
K.~Mejrhit and R.~Hajji,
Eur. Phys. J. C \textbf{80}, no.11, 1060 (2020)
doi:10.1140/epjc/s10052-020-08632-1

\bibitem{Nakarachinda:2021jxd}
R.~Nakarachinda, E.~Hirunsirisawat, L.~Tannukij and P.~Wongjun,
Phys. Rev. D \textbf{104}, no.6, 064003 (2021)
doi:10.1103/PhysRevD.104.064003
[arXiv:2106.02838 [gr-qc]].

\bibitem{Abreu:2021avp}
E.~M.~C.~Abreu and J.~Ananias Neto,
EPL \textbf{133}, no.4, 49001 (2021)
doi:10.1209/0295-5075/133/49001

\bibitem{Renyi60} A. R\'enyi, ``On measures of information and 
entropy'', in Proceedings of the fourth Berkeley Symposium on 
Mathematics, Statistics and Probability,
20~June-30 July 1960, Volume~I, University of California Press, Berkeley 
and Los Angeles, pp. 547-561.

\bibitem{Tsallis:1987eu}
C.~Tsallis,
J. Statist. Phys. \textbf{52}, 479-487 (1988)
doi:10.1007/BF01016429

\bibitem{Szabados:2009eka}
L.~B.~Szabados,
Living Rev. Rel. \textbf{12}, 4 (2009)
doi:10.12942/lrr-2009-4

\bibitem{Misner:1964je}
C.~W.~Misner and D.~H.~Sharp,
Phys. Rev. \textbf{136}, B571-B576 (1964)
doi:10.1103/PhysRev.136.B571

\bibitem{Hernandez:1966zia} W.~C.~Hernandez and C.~W.~Misner,
Astrophys. J. \textbf{143} (1966), 452 doi:10.1086/148525

\bibitem{Brown:1992br}
J.~D.~Brown and J.~W.~York, Jr.,
Phys. Rev. D \textbf{47}, 1407-1419 (1993)
doi:10.1103/PhysRevD.47.1407
[arXiv:gr-qc/9209012 [gr-qc]].

\bibitem{McVittie} G.~C. McVittie, Mon. Not. Roy. Astr. Soc. {\bf 93}, 325 
(1933).

\bibitem{Carrera:2009ve}
M.~Carrera and D.~Giulini,
Phys. Rev. D \textbf{81}, 043521 (2010)
doi:10.1103/PhysRevD.81.043521
[arXiv:0908.3101 [gr-qc]].

\bibitem{Faraoni:2015ula}
V.~Faraoni, {\it Cosmological and Black Hole Apparent Horizons} 
Lect. Notes Phys. \textbf{907} (Springer, New York, 2015) 
doi:10.1007/978-3-319-19240-6

\bibitem{SultanaDyer05} J. Sultana and C.C. Dyer, 
{\it Gen. Relativ. Gravit.} {\bf 37}, 1347  
(2005).

\bibitem{Hayward:1994bu}
S.~A.~Hayward,
Phys. Rev. D \textbf{53}, 1938-1949 (1996)
doi:10.1103/PhysRevD.53.1938
[arXiv:gr-qc/9408002 [gr-qc]].

\bibitem{Hayward:1998ee}
S.~A.~Hayward, S.~Mukohyama and M.~C.~Ashworth,
Phys. Lett. A \textbf{256}, 347-350 (1999)
doi:10.1016/S0375-9601(99)00225-X
[arXiv:gr-qc/9810006 [gr-qc]].

\bibitem{Hayward:1997jp}
S.~A.~Hayward,
Class. Quant. Grav. \textbf{15}, 3147-3162 (1998)
doi:10.1088/0264-9381/15/10/017
[arXiv:gr-qc/9710089 [gr-qc]].

\bibitem{Wald:1993nt}
R.~M.~Wald,
Phys. Rev. D \textbf{48}, no.8, R3427-R3431 (1993)
doi:10.1103/PhysRevD.48.R3427
[arXiv:gr-qc/9307038 [gr-qc]].

\bibitem{Iyer:1994ys}
V.~Iyer and R.~M.~Wald,
Phys. Rev. D \textbf{50}, 846-864 (1994)
doi:10.1103/PhysRevD.50.846
[arXiv:gr-qc/9403028 [gr-qc]].

\bibitem{Kodama:1979vn}
H.~Kodama,
Prog. Theor. Phys. \textbf{63}, 1217 (1980)
doi:10.1143/PTP.63.1217.

\bibitem{Vanzo:2011wq}
L.~Vanzo, G.~Acquaviva and R.~Di Criscienzo,
Class. Quant. Grav. \textbf{28}, 183001 (2011)
doi:10.1088/0264-9381/28/18/183001
[arXiv:1106.4153 [gr-qc]].

\bibitem{Nojiri:2019skr}
S.~Nojiri, S.~D.~Odintsov and E.~N.~Saridakis,
Eur. Phys. J. C \textbf{79} (2019) no.3, 242
doi:10.1140/epjc/s10052-019-6740-5
[arXiv:1903.03098 [gr-qc]].

\bibitem{Barrabes:1997kk}
C.~Barrabes and G.~F.~Bressange,
Class. Quant. Grav. \textbf{14}, 805-824 (1997)
doi:10.1088/0264-9381/14/3/021
[arXiv:gr-qc/9701026 [gr-qc]].

\bibitem{Bressange:2000qp}
G.~F.~Bressange,
Class. Quant. Grav. \textbf{17}, 2509-2524 (2000)
doi:10.1088/0264-9381/17/13/304
[arXiv:gr-qc/0005021 [gr-qc]].

\bibitem{Velay-Vitow:2017odc}
J.~Velay-Vitow and A.~DeBenedictis,
Phys. Rev. D \textbf{96}, no.2, 024055 (2017)
doi:10.1103/PhysRevD.96.024055
[arXiv:1705.02533 [gr-qc]].

\bibitem{Berezin:2019maf}
V.~Berezin, V.~Dokuchaev, Y.~Eroshenko and A.~Smirnov,
Int. J. Mod. Phys. A \textbf{35}, no.02n03, 2040002 (2020)
doi:10.1142/S0217751X20400023
[arXiv:1909.06405 [gr-qc]].

\bibitem{Berezin:2020mas}
V.~A.~Berezin, V.~I.~Dokuchaev, Y.~N.~Eroshenko and A.~L.~Smirnov,
Class. Quant. Grav. \textbf{38}, no.4, 045014 (2021)
doi:10.1088/1361-6382/abd143
[arXiv:2008.01813 [gr-qc]].

\bibitem{Berezin:2020bvs}
V.~A.~Berezin, V.~I.~Dokuchaev, Y.~N.~Eroshenko and A.~L.~Smirnov,	
Phys. Part. Nucl. \textbf{51}, no.4, 730-734 (2020)
doi:10.1134/S1063779620040139
[arXiv:2009.12400 [gr-qc]].

\bibitem{Aviles:2019xae}
L.~Avil\'es, H.~Maeda and C.~Martinez,
Class. Quant. Grav. \textbf{37}, no.7, 075022 (2020)
doi:10.1088/1361-6382/ab728a
[arXiv:1910.07534 [gr-qc]].
 
\bibitem{Olmo:2020fri}
G.~J.~Olmo and D.~Rubiera-Garcia,
Class. Quant. Grav. \textbf{37}, no.21, 215002 (2020)
doi:10.1088/1361-6382/abb924
[arXiv:2007.04065 [gr-qc]].

\bibitem{Chu:2021uec}
C.~S.~Chu and H.~S.~Tan,
[arXiv:2103.06314 [hep-th]].

\bibitem{Rosa:2021teg}
J.~L.~Rosa,
Phys. Rev. D \textbf{103}, no.10, 104069 (2021)
doi:10.1103/PhysRevD.103.104069
[arXiv:2103.11698 [gr-qc]].

\bibitem{Brans:1961sx} C.~Brans and R.~H.~Dicke,
Phys. Rev. \textbf{124} (1961), 925-935 doi:10.1103/PhysRev.124.925

\bibitem{Sotiriou:2008rp} T.~P.~Sotiriou and V.~Faraoni,
Rev. Mod. Phys. \textbf{82} (2010), 451-497 doi:10.1103/RevModPhys.82.451 
[arXiv:0805.1726 [gr-qc]].

\bibitem{Nojiri:2017ncd}
S.~Nojiri, S.~D.~Odintsov and V.~K.~Oikonomou,
Phys. Rept. \textbf{692} (2017), 1-104
doi:10.1016/j.physrep.2017.06.001
[arXiv:1705.11098 [gr-qc]].

\bibitem{Nojiri:2010wj} S.~Nojiri and S.~D.~Odintsov,
Phys. Rept. \textbf{505} (2011), 59-144 doi:10.1016/j.physrep.2011.04.001 
[arXiv:1011.0544 [gr-qc]].

\bibitem{Hawking:1972qk} S.~W.~Hawking,
Commun. Math. Phys. \textbf{25} (1972), 167-171 doi:10.1007/BF01877518

\bibitem{Bekenstein:1996pn} J.~D.~Bekenstein,
[arXiv:gr-qc/9605059 [gr-qc]].

\bibitem{Sotiriou:2011dz} T.~P.~Sotiriou and V.~Faraoni,
Phys. Rev. Lett. \textbf{108} (2012), 081103 
doi:10.1103/PhysRevLett.108.081103 [arXiv:1109.6324 [gr-qc]].

\bibitem{Bhattacharya:2015iha} S.~Bhattacharya, K.~F.~Dialektopoulos, 
A.~E.~Romano and T.~N.~Tomaras,
Phys. Rev. Lett. \textbf{115} (2015) no.18, 181104 
doi:10.1103/PhysRevLett.115.181104 [arXiv:1505.02375 [gr-qc]].

\bibitem{Barrow:2020tzx}
J.~D.~Barrow,
Phys. Lett. B \textbf{808} (2020), 135643
doi:10.1016/j.physletb.2020.135643
[arXiv:2004.09444 [gr-qc]].

\bibitem{Cognola:2005de}
G.~Cognola, E.~Elizalde, S.~Nojiri, S.~D.~Odintsov and S.~Zerbini,
JCAP \textbf{02} (2005), 010
doi:10.1088/1475-7516/2005/02/010
[arXiv:hep-th/0501096 [hep-th]].

\bibitem{Volovik:2021upi}
G.~E.~Volovik,
[arXiv:2107.11193 [gr-qc]].

\bibitem{Gibbons:1977mu}
G.~W.~Gibbons and S.~W.~Hawking,
Phys. Rev. D \textbf{15}, 2738-2751 (1977)
doi:10.1103/PhysRevD.15.2738

\bibitem{Babichev:2004yx}
E.~Babichev, V.~Dokuchaev and Y.~Eroshenko,
Phys. Rev. Lett. \textbf{93}, 021102 (2004)
doi:10.1103/PhysRevLett.93.021102
[arXiv:gr-qc/0402089 [gr-qc]].


\end{thebibliography}
\end{document}